\documentclass[prl,aps,twocolumn,showpacs,superscriptaddress]{revtex4}

\usepackage{amsfonts}

\usepackage{graphicx}

\begin{document}

\title{High-magnetic field lattice length changes in URu$_{2}$Si$_2$}

\author{V. F. Correa}
\affiliation{Centro At\'omico Bariloche (CNEA) and Instituto Balseiro (U. N. Cuyo), 8400 Bariloche, R\'io Negro, Argentina}

\author{S. Francoual}
\altaffiliation[Present address: ]{Deutsches Elektronen-Synchrotron DESY, D-22607 Hamburg, Germany}
\affiliation{MPA-CMMS, Los Alamos National Laboratory, Los Alamos, New Mexico 87545, USA}

\author{M. Jaime}
\affiliation{MPA-CMMS, Los Alamos National Laboratory, Los Alamos, New Mexico 87545, USA}

\author{N. Harrison}
\affiliation{MPA-CMMS, Los Alamos National Laboratory, Los Alamos, New Mexico 87545, USA}

\author{T. P. Murphy}
\affiliation{National High Magnetic Field Laboratory, Florida State University, Tallahassee, Florida 32310, USA}

\author{E. C. Palm}
\affiliation{National High Magnetic Field Laboratory, Florida State University, Tallahassee, Florida 32310, USA}

\author{S. W. Tozer}
\affiliation{National High Magnetic Field Laboratory, Florida State University, Tallahassee, Florida 32310, USA}

\author{A. H. Lacerda}
\affiliation{LANSCE, Los Alamos National Laboratory, Los Alamos, New Mexico 87545, USA}

\author{P. A. Sharma}
\altaffiliation[Present address: ]{Sandia National Laboratories, Albuquerque, NM 87185, USA}
\affiliation{MPA-CMMS, Los Alamos National Laboratory, Los Alamos, New Mexico 87545, USA}

\author{J. A. Mydosh}
\affiliation{Kamerlingh Onnes Laboratory, Leiden University, NL-2300 RA Leiden, The Netherlands}

\date{\today}
\pacs{71.27.+a, 65.40.De, 75.80.+q, 75.30.Kz}

\begin{abstract}

We report high magnetic field (up to 45 T) $\hat{c}$-axis thermal expansion and magnetostriction experiments on URu$_{2}$Si$_2$ single crystals. 
The sample length change $\Delta L_c(T_{HO})  / L_c$ associated with the transition to the ``hidden order'' phase becomes increasingly discontinous as the magnetic field is raised above 25 T. The re-entrant ordered phase III is clearly observed in both the thermal expansion $\Delta L_c(T)  / L_c$ and magnetostriction $\Delta L_c(B)  / L_c$ above 36 T, in good agreement with previous results. The sample length is also discontinuous at the boundaries of this phase, mainly at the upper boundary. 
A change in the sign of the coefficient of thermal-expansion $\alpha_c = \frac{1}{L_c}\left(\frac {\partial \Delta L_c}{\partial T} \right)$ is observed at the metamagnetic transition ($B_M$ $\sim$ 38 T) which is likely related to the existence of a quantum critical end point.

\end{abstract}

\maketitle

Even though the 4$f^1$ Ce-based compounds \cite{Lohneysen} are by far the most extensively studied heavy fermions, it is a 5$f$ uranium compound, namely URu$_2$Si$_2$, which remains as one of the most intriguing and unsolved systems among the strongly correlated electron materials \cite{Mydosh}.
With an effective mass $m^* \sim$ 25 - 50 $m_e$, it is considered a moderate heavy fermion \cite{Palstra,Maple,Schlabitz}.
URu$_2$Si$_2$ becomes an unconventional superconductor at $T_c$ = 1.2 K. Well above $T_c$, a clear second order (mean field-like) transition occurs at $T_{HO} \approx$ 17 K originally identified as the onset of antiferromagnetism. However, the tiny (0.03 $\mu_B$) ordered moment observed through neutron diffraction experiments \cite{Broholm,Mason} is unable to account for the large entropy change ($\sim 0.15 - 0.25 R$) at $T_{HO}$ \cite{Palstra,Maple}. It is currently accepted that this small staggered moment is parasitic to this ``hidden order'' phase \cite{Niklowitz,Baek}. Hydrostatic pressure, however, is detrimental to both superconductiviy and HO favoring instead large-moment antiferromagnetism \cite{Amitsuka, Hassinger}. Progressive doping with small amounts of Rh replacing Ru also supresses the HO phase \cite{Yokoyama}.

Under magnetic field, URu$_2$Si$_2$ exhibits a fascinating behavior. As usual, superconductivity is supressed by just a few Tesla. Nevertheless, about 35 T are needed to destroy the HO phase. Above 25 T, specific heat experiments show that this transition becomes very sharp and narrow losing its second order character \cite{Jaime}. 
Different experiments reveal several transitions or crossovers at higher fields resulting in an intricate temperature versus magnetic field phase diagram \cite{Jaime,KHK,Scheerer}. 
A possible re-entrant order (phase III) is observed between 36 and 39 T, which terminates at a polarized Fermi liquid at the high field limit \cite{Jaime,KHK,Scheerer}.
At temperatures above this dome-shaped phase, the magnetization along the $\hat{c}$-axis (easy axis) increases non monotononically as the field is raised showing a maximum change ($dM/dB$) at $B_M$ $\sim$ 38 T \cite{Harrison}. As the temperature is lowered this inflection point becomes more pronounced and it would acquire an infinite slope at $T$ = 0 (becoming a true quantum phase transition) provided the field induced phases were absent. 
This picture is supported by resistivity experiments \cite{KHK} that show a collapse of the effective Fermi temperature $T^*$ at $B_M$, thus indicating that this characteristic field might correspond to a quantum critical endpoint (QCEP) as is observed in CeRu$_2$Si$_2$ \cite{Sugi} and Sr$_3$Ru$_2$O$_7$ \cite{Grigera}. 
On the other hand, the presence of the high-field phases results in two distinctive magnetization plateaus \cite{Scheerer}.
 
Various theories and models were proposed to explain the HO pressure and doping dependence and to unveil the nature of the mysterious HO phase \cite{Mydosh}. Roughly, models can be ascribed to two different groups: U-5$f$ electrons treated as localized electrons or contributing to the Fermi surface as itinerant electrons. The diverse proposed HO order parameters, however, cover an amazingly wide spectrum: orbital, multipolar, spin nematic, hybridization and spin-inter-orbital density waves, modulated spin liquid, hastatic and dynamical.
Yet, little attention has been given to the interpretation of the high-field phases \cite{Mydosh}. 

The renewed interest in URu$_2$Si$_2$ has been triggered in recent years by a new set of high quality experiments using microscopic \cite{Aynaijian,Schmidt}, spectroscopic \cite{Wiebe,Janik,Santander,Walker,Kawasaki,Dakovski,Liu,Levallois,Guo} and transport experiments \cite{Jo,Shishido,Altarawneh,Malone,Scheerer}. It has become more or less clear that the HO is accompanied by a partial gapping of the Fermi surface at certain momentum hot spots, that incommensurate and commensurate itinerant magnetic excitations are present in the HO phase, and that nesting induces a reconstruction of the Fermi surface in the HO phase.
Macroscopic techniques, on the other hand, have mainly characterized the clear second order phase transition at 17 K. Nevertheless, thermodynamic studies are far from complete and they can still give valuable information. The high magnetic field properties and their connection to the HO are, for instance, an ongoing study \cite{Scheerer}.

In this work we report high static magnetic field linear thermal expansion and magnetostriction results on URu$_2$Si$_2$ single crystals. At low $B$ ($<$ 25 T) a typical second order-like feature characterizes the transition to the HO phase in good agreement with previous results \cite{deVisser,Mentink}. The transition, however, becomes increasingly sharper as $B$ is raised above 25 T to culminate in a clear discontinous first order-like transition before its suppression at 35 T. A new phase emerges between 36 and 39 T giving rise to a dome-like region (phase III) in the $T-B$ phase diagram (see Fig.~\ref{fig4}). The sample length change at the dome borders is discontinous (to a greater extent at the high-field border) but evolves to a continous transition at the top of the dome. Around $B =$ 38 T ($\approx$ top of the dome), the low-$T$ (above the ordered phases) thermal-expansion coefficient reverses sign in a manner that is consistent with an entropy accumulation around an underlying QCEP \cite{Garst}.

Single crystals of URu$_{2}$Si$_2$ were grown by the Czochralski method. The selected sample was cut in a cubic shape of length $L\approx$ 3 mm.
A capacitive dilatometer \cite{Schmiedeshoff} was used in our experiments, achieving a resolution of 5$\cdot$10$^{-7}$ in DC fields to 45 T. The setup is placed in an environment with a low pressure ($P<$ 10$^{-1}$ torr) of $^3$He gas. Experiments were performed in the hybrid 45 T magnet at the NHMFL, Tallahassee. All the results presented here were obtained in the longitudinal configuration, i.e., $B \parallel$ tetragonal $\hat{c}$-axis $\parallel L$. 

\begin{figure}[]
\includegraphics[width=\columnwidth]{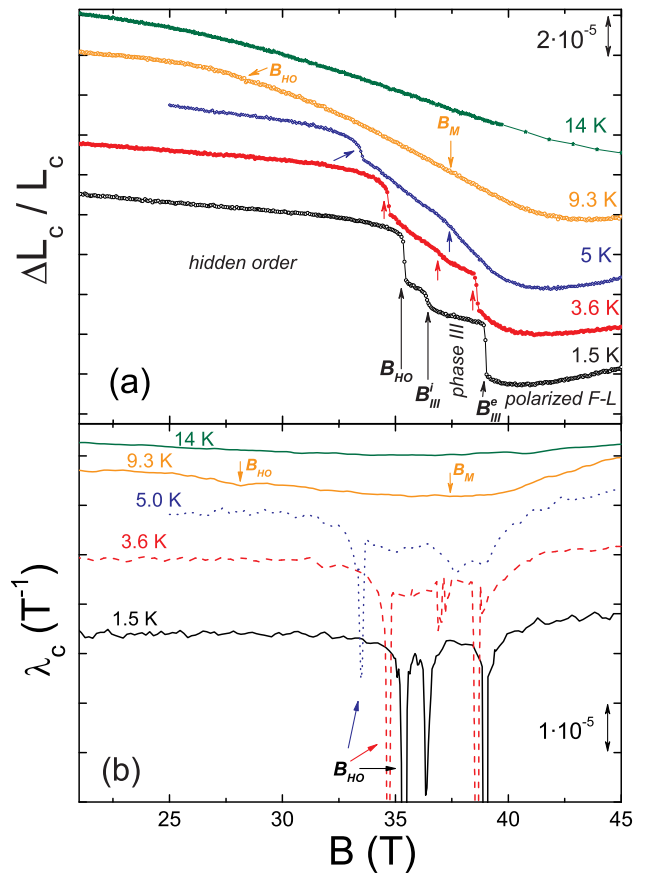}
\caption[]{(color online). (a) Field dependence of the $\hat{c}$-axis magnetostriction at different temperatures and (b) its derivative, the magnetostriction coefficient $\lambda_c$. Arrows indicate the different transitions, labeled as described in the main text. Curves are vertically shifted.}
\label{fig1}
\end{figure}

The isothermal linear magnetostriction $\Delta L_c(B)  / L_c$ at different temperatures is displayed in Fig.~\ref{fig1}(a). The different transitions are labeled and indicated by arrows. The curve measured at 1.5 K shows the different field induced transitions: $B_{HO}$, $B^i_{III}$ and $B^e_{III}$ denoting, respectively, the HO transition, the entry to and the exit of phase III. The sharp and narrow shape of the transitions (mainly $B_{HO}$ and $B^e_{III}$) reflects their discontinous first order-like character. 
Note that (i) the sample length changes at the three transitions are negative and that (ii) the magnitude of these jumps at $B_{HO}$ and $B^e_{III}$ are very similar. 
As $T$ is raised, however, the length changes at the transitions become smaller and less abrupt. This behavior, however, is not symmetric: the first order-like character at $B^e_{III}$ seems to extend to higher temperatures than that at $B^i_{III}$, as sketched in Fig.~\ref{fig4}. Finally, $B^i_{III}$ and $B^e_{III}$ merge in a single wide kink corresponding to the top of phase III dome (see curve at 5 K). 
The progressive reduction of discontinous character is even more evident at the HO transition, $B_{HO}$. Above 10 K, in fact, $B_{HO}$ is virtually imperceptible.
The corresponding magnetostriction coefficient $\lambda_c = \frac{1}{L_c}\left(\frac {\partial \Delta L_c}{\partial B} \right)$ is displayed in Fig.~\ref{fig1}(b).
The evolution of $B_{III}$ from a discontinous to a continous transition when increasing $T$ (i.e., when going from the bottom to the top of the dome) is evident.
Above the dome ($T >$ 5 K) a tiny and broad bump persists at higher temperatures as seen in the curve at 9.3 K. This subtle dip may be linked to the metamagnetic transition underneath phase III dome \cite{Harrison}.

\begin{figure}[]
\includegraphics[width=\columnwidth]{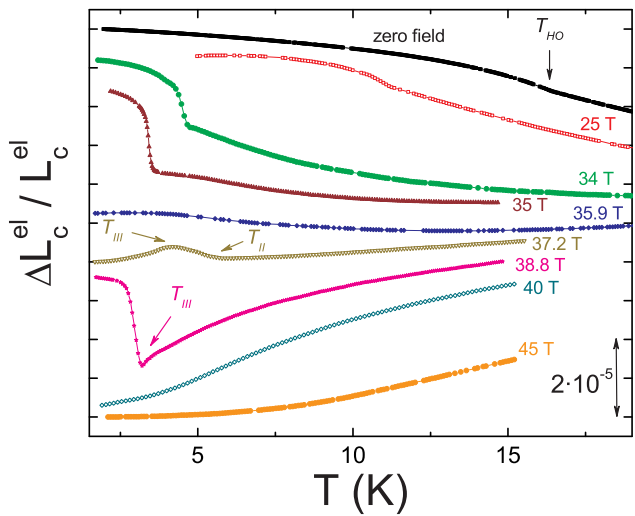}
\caption[]{(color online). Electronic contribution to $\hat{c}$-axis expansivity at different magnetic fields. Arrows indicate the different transitions, labeled as described in the main text. Curves are vertically shifted.}
\label{fig2}
\end{figure}

Figure~\ref{fig2} shows the electronic contribution to the expansivity $\Delta L^{el}_c(T)  / L^{el}_c$ at different $B$. Lattice-phonon contribution has been substracted using a numerically-generated Debye curve ($\theta_D$ = 300 K) \cite{Palstra,Maple} since no data from a related compound without $f$ electrons (e.g., ThRu$_{2}$Si$_2$) are presently available. The calculated phonon contribution matches well the zero field expansivity above 45 K giving us confidence in the adopted procedure. 
Both, the HO transition $T_{HO}$ and phase III transition $T_{III}$ can be observed. The evolution of $T_{HO}$ from a second order to a first order-like transition as $B$ is raised is clearly seen. The HO phase is suppressed for $B >$ 35 T. Phase III is then observed between 36 and 39 T.

\begin{figure}[th]
\includegraphics[width=\columnwidth]{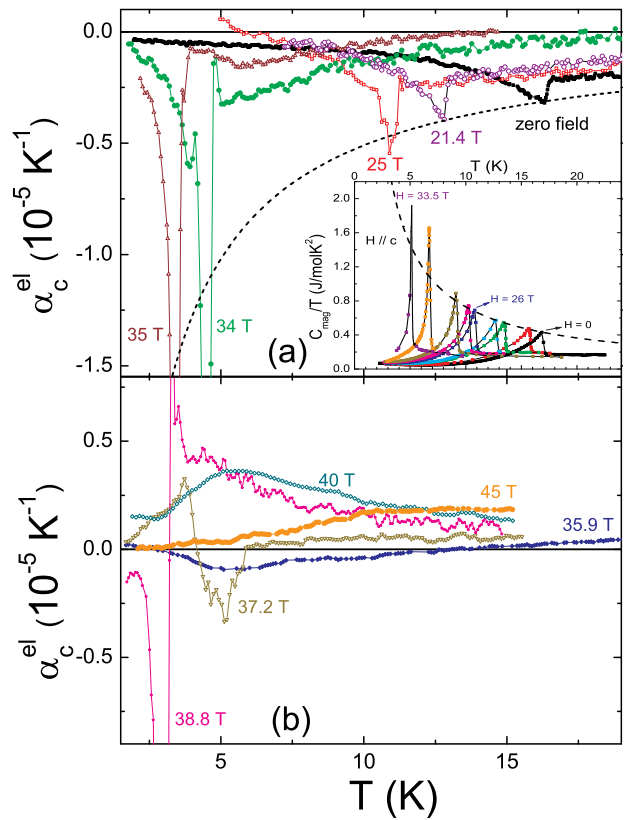}
\caption[]{(color online). Electronic contribution to the $\hat{c}$-axis thermal-expansion coefficient for (a) $B <$ 35 T and (b) $B >$ 35 T. Inset: Electronic contribution to the specific heat for $B <$ 35 T. Dashed lines correspond to a $1/T$ dependence (see text).}
\label{fig3}
\end{figure}

\begin{figure}[]
\includegraphics[width=\columnwidth]{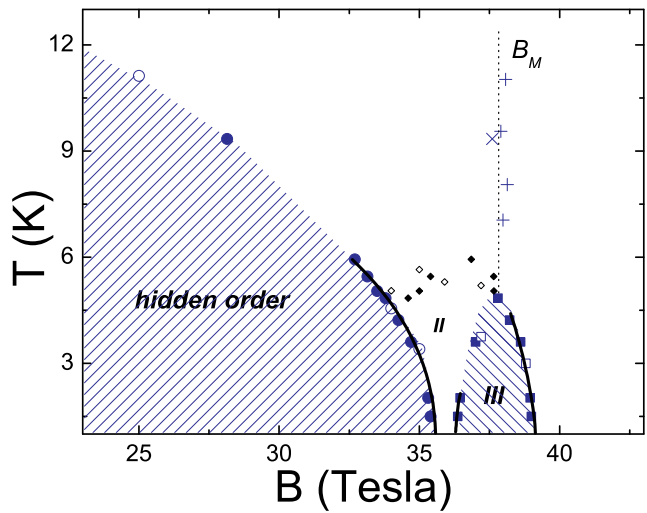}
\caption[]{(color online). Temperature - magnetic field phase diagram. Open (closed) symbols extracted from thermal expansion (magnetostriction) experiments. $\times$ (+) denotes the metamagnetic transition $B_M$ from this work (from Ref. [\onlinecite{Harrison}]). Bold solid lines correspond to clear first order phase transitions.}
\label{fig4}
\end{figure}

It is noteworthy to analyze the temperature dependence of $\Delta L^{el}_c(T)  / L^{el}_c$ above the ordered phases. This is better achieved through the thermal-expansion coefficient $\alpha^{el}_c = \frac{1}{L^{el}_c}\left(\frac {\partial \Delta L^{el}_c}{\partial T} \right)$ as shown in Fig.~\ref{fig3}. 
Above $T_{HO}$, $\alpha^{el}_c$ is negative as seen in Fig.~\ref{fig3}(a). At higher fields ($B >$ 35 T), however, Fig.~\ref{fig3}(b) shows that $\alpha^{el}_c$ becomes positive. 
A rather similar sign change has been observed previously in CeRu$_2$Si$_2$ \cite{Paulsen} and Sr$_3$Ru$_2$O$_7$ \cite{Gegenwart}, 
and it was associated with a non-symmetry breaking metamagnetic quantum critical endpoint (QCEP) \cite{Garst,Gegenwart,Weickert}.
The similarity is notable despite the fact that the microscopic origin of the QCEP may be rather different in the three systems.
In URu$_{2}$Si$_2$ the putative QCEP is masked by a phase dome as in Sr$_3$Ru$_2$O$_7$. 

Figure~\ref{fig4} shows a $T-B$ phase diagram summarizing our observations. Solid symbols are extracted from magnetostriction experiments while open symbols are taken from thermal-expansion experiments. We have included the metamagnetic fields from Ref.[\onlinecite{Harrison}]. Small diamond symbols account for relatively small features observed in both experiments (see curve at 37.2 T in Fig.~\ref{fig2} for instance) which roughly trace previously reported phase II \cite{KHK,Suslov}. 

Our high-field lattice expansion experiments clearly show that the HO transition becomes first order above $\approx$ 25 T. This is demonstrated by the asymmetric lambda-like transitions at low fields that become sharp negative spikes in $\alpha_c$ and $\lambda_c$ as seen in Figs.~\ref{fig3} and ~\ref{fig1}, respectively. 
In fact, if we connect the HO peak in $\alpha_c(T)$ along with that of $C/T$ \cite{Jaime}, we note a nice $1/T$ dependence from 0 to 25 T, as shown by the dashed lines in Fig.~\ref{fig3} and its inset. Above 25 T, the peak gradually starts to exceed the $1/T$ behavior indicating that it is losing its second order character.
Entrance to the low ($B^i_{III}$) and high ($B^e_{III}$) field sides of the phase III dome (at 36 and 39 T, respectively) are also first order transitions at low temperature, here again represented by similar negative spikes at $B^i_{III}$ ($B^e_{III}$) at and below 2.1 K (4.2 K). By crossing the dome at its top $\lambda_c$ exhibits only a small dip (Fig.~\ref{fig1}) while $\alpha_c$ exhibits a change of sign (Fig.~\ref{fig3}) possibly associated with an underlying QCEP, masked by the novel phase III. This feature in the data points to a continous (second order) phase transition at the  dome's top.

As shown in Fig.~\ref{fig2} there is a clear change in the temperature dependence of $\Delta L_c(T)  / L_c$ between 35.9 and 37.2 T. These fields are similar to the second plateau fields, between 36.3 and 37.4 T recently observed in the magnetization \cite{Scheerer}, and the decrease in entropy previously found in the magnetocaloric effect at 36 T \cite{Jaime}. By combining these results we can describe phase III as a contracted $\hat{c}$-axis, weakly magnetically polarized $\hat{c}$-axis ordered phase. Here the large field polarizes the Fermi surfaces so as to destroy the hybridization between the conduction electrons and the 5$f$-$U$ ions, thereby allowing a partial magnetic moment to form along the field direction. Such is known in the present literature as Kondo breakdown \cite{Pepin}.

In summary, our $\hat{c}$-axis length changes define three first order phase transitions as $T \rightarrow$ 0: the field quenching of HO at 35 T and the entry and exit fields of phase III. The previous detection of the metamagnetic transition above the dome \cite{Harrison} with magnetization vs. field data and the change of sign in the present thermal expansion study suggest an underlying quantum critical end point with a possible second order quantum phase transition (if QCEP is located at $T$ = 0 K) now hidden by the dome - a common phenomenon in heavy-fermion quantum critical behavior. The change of sign $\alpha_c$, where $\alpha_c \propto \left( \partial S / \partial H \right)_T$, points to the existence of a maximum in the entropy as a function of field at $B_M$, and is believed to be due to soft quantum fluctuations surrounding the QCEP \cite{Garst}.

The exact nature of the novel phase III remains unclear. Yet, a comparison with Sr$_3$Ru$_2$O$_7$ may be worthwhile. The metamagnetic QCEP in Sr$_3$Ru$_2$O$_7$ \cite{Grigera} is masked by a dome associated with an electronic nematic fluid phase \cite{Borzi}. The borders of that dome correspond to first order phase transitions while the top is characterized by a continous transition \cite{Grigera2}. The sample length change at the entry (low field) and exit (high field) of the dome are of the same sign \cite{Grigera}. All these features are common to phase III in URu$_{2}$Si$_2$. So, an interesting possibility is to relate this novel phase to the magnetic signature of an electronic nematic phase. 
Close proximity to other phases (HO and mainly phase II) adds complexity to the scenario in URu$_{2}$Si$_2$. 
Here additional theory work is required to properly explain this novel phase. 

Because of the very large static fields required and the occurrence of multiple phases extended to relative high tempeatures, our thermal expansion data are insufficient to attempt a detailed thermodynamic analysis of the metamagnetic QCEP as was done for non-phase transitioned CeRu$_{2}$Si$_2$ \cite{Weickert} and low field/temperature transitioned Sr$_3$Ru$_2$O$_7$ \cite{Gegenwart}. We also need new angular dependent measurements away from the $\hat{c}$-axis into the $\hat{a}$-$\hat{a}$ plane to test a possible nematic transition in phase III as suggested for Sr$_3$Ru$_2$O$_7$. Such experiments are planned for future studies of lattice parameter changes in the 40 T region.

Work at the NHMFL was performed under the auspices of the National Science Foundation (NSF Cooperative Agreement No. DMR-0654118), the State of Florida and the DOE/NNSA under DE-FG52-10NA29659. Work partially supported by ANPCyT PICT05-32900, Argentina. V. F. C. is a member of CONICET, Argentina. S. F. acknowledges fundship from the Seaborg Institute and from NSF.


\end{document}